\documentclass[preprint,english,aps,showpacs]{revtex4-1}
\pdfoutput=1
\usepackage{amsmath}
\usepackage{color}
\usepackage{graphicx}
\usepackage{epstopdf}
\usepackage{babel}

\begin{document}

\title{Structure of Fe$_3$Si/Al/Fe$_3$Si thin film stacks on GaAs(001)}
\author{B.~Jenichen}
\email{bernd.jenichen@pdi-berlin.de}
\author{U.~Jahn}
\author{A.~Nikulin}

\author{J.~Herfort}

\affiliation{Paul-Drude-Institut f\"ur Festk\"orperelektronik,
Hausvogteiplatz 5--7,D-10117 Berlin, Germany}

\author{H.~Kirmse}
\affiliation{Humboldt University of Berlin, Institute of Physics,
Newtonstrasse 15,
D-12489 Berlin,  Germany}

\date{\today}

\begin{abstract}
Fe$_{3}$Si/Al/Fe$_{3}$Si/GaAs(001) structures were deposited by molecular-beam epitaxy and characterized by transmission and scanning electron microscopy, and  x-ray diffraction. The first Fe$_{3}$Si  film on GaAs(001) is growing epitaxially as (001) oriented single crystal. The subsequent Al film grows almost \{111\} oriented in a fibre texture although the underlying Fe$_{3}$Si is exactly (001) oriented. The growth in this orientation is triggered by a thin transition region which is formed at the Fe$_{3}$Si/Al interface. In the end after the growth of the second Fe$_{3}$Si layer on top of the Al the final properties of the whole stack depend on the substrate temperature T$_s$ during deposition of the last film. The upper Fe$_{3}$Si films are mainly \{110\} oriented although they are poly-crystalline. At lower T$_s$, around room temperature, all the films retain their original structural properties.
\end{abstract}

\pacs{68.70.+w, 68.55.ag, 68.37.Lp, 61.05.cp}

\maketitle

\section{Introduction}
Device concepts based on the spin rather than the charge of the electron have been
explored recently. These concepts will lead to further improvements in device performance.\cite{Zutic2004,Palmstrom2003} Nonvolatile memory technology using nanopillar magnetic tunnel junctions (MTJ) with spin transfer torque (STT) switching is currently most promising.\cite{Ikeda2007, Slonczewski1995} In general, a STT device consists of two ferromagnetic (FM) films separated by a thin spacer layer, which can be either on oxide barrier or a non-FM metal.\cite{Ralph2008, Slonczewski1989} Via STT the orientation of a thin free magnetic layer in a MTJ can be changed using a spin-polarized current pulse, which is created in the thicker fixed magnetic layer. Hence, it can be used to flip the active elements in a magnetic random-access memory (RAM). Such STT-RAM has a low power consumption and good scalability in comparison to a magneto-resistive RAM.

In addition, particularly because of their potential for non-volatile memory applications, hybrid structures consisting of ferromagnetic (FM) metals and semiconductors (SC) are of major interest for the field of semiconductor spintronics. Here the FM acts as an injection layer for creating spinpolarized carriers within the SC. In general, a single FM film on top of a SC layer is sufficient for this purpose. We have recently demonstrated all-electrical spin injection and detection in local, non-local as well as extraction spin valves based on the Co$_{2}$FeSi/GaAs hybrid system \cite{Bruski2013,Manzke2013}, which are essential building blocks for spintronic applications. In such spin valve structures the switching of the magnetization within the ferromagnetic contacts is obtained by applying an external magnetic field. The magnitude of the external magnetic field required for the reversal of the magnetization slightly differs for each contact due to small geometrical differences. The integration of STT structures in SC spintronic devices such as spin valves would on the other hand allow for an electrical switching of individually addressed FM contacts.

In the present work we grow and analyze ferromagnet/metal/ferromagnet (F/M/F) structures: Fe$_{3}$Si/Al/Fe$_{3}$Si layer stacks grown on GaAs(001) by molecular beam epitaxy (MBE). The misfit of stoichiometric Fe$_{3}$Si and GaAs is very low.\cite{herfort03} The as-grown structures were characterized by transmission electron microscopy (TEM), energy dispersive x-ray spectroscopy (EDX) in the TEM, scanning electron microscopy (SEM) mainly in the electron backscattered diffraction (EBSD) mode, and x-ray diffraction (XRD) combined with phase retrieval.






\section{Experiment}
The structures were grown by MBE on GaAs (001) substrates. The main growth parameters are given in Table~\ref{tab:tab1}. After growth of a GaAs 300~nm thick buffer layer (T$_s$ = 580~$^\circ$C, growth rate 480~nm/h) the templates were transferred under ultra high vacuum conditions to the As free growth chamber of the MBE system. Then nominally 4.5-nm-thick Fe$_{3}$Si films were grown at a substrate temperature of 200~$^\circ$C as described previously.\cite{herfort03} Here, 200~$^\circ$C is within the optimum growth temperature range to obtain ferromagnetic Fe$_{3}$Si layers with high crystalline and interfacial perfection on GaAs(001).\cite{herfort2005,herfort2006,herfort2006PE}

 Aluminium films were grown on top of these Fe$_{3}$Si films  at a substrate temperature of 0~$^\circ$C. The growth rates were chosen to be 18~nm/h and 285~nm/h for Fe$_{3}$Si and Al respectively. On top of the Al film a second Fe$_{3}$Si film was grown at two different substrate temperatures: one sample at T$_s$ = 17-67~$^\circ$C (temperature ramp) and the second sample in the beginning at T$_s$ = 13-25~$^\circ$C (temperature ramp) and later at T$_s$ = 200~$^\circ$C. The temperature ramp occurs because of the temperature drift during the growth time of 150~min due to the heat radiation of the source without additional cooling oft the substrate.  Another sample was left without the second Fe$_{3}$Si film for comparison.

First the samples were investigated by high-resolution (HR) XRD  in order to obtain macroscopic information about the structure of the upper Fe$_{3}$Si and the Al films. XRD measurements were performed on the structures using a Panalytical X-Pert PRO MRD\texttrademark\ system with a Ge(220) hybrid monochromator  (Cu~K$\alpha_1$ radiation with a wavelength of $\lambda=1.54056$~\AA). The program Epitaxy\texttrademark\ was used for data evaluation. For the direct determination of the displacement depth profile from the XRD curves we used the x-ray phase retrieval method.~\cite{nikulin1998,nikulin1999}

In addition the as-grown structures were characterized on a microscale  by HR~TEM.  For that purpose cross-sectional TEM
specimens were prepared by mechanical lapping and polishing, followed by argon ion milling according to standard techniques.
TEM images were acquired with a JEOL 3010 microscope operating at 200~kV in order to reduce radiation damage. The cross section TEM method provides
high lateral and depth resolutions on the nanometer scale, however, they average over the thickness of the thin sample foil ($\sim$~20~nm).

Elemental analyses were performed by high-resolution scanning transmission electron microscope (STEM). STEM EDX mapping was obtained using the JEOL 2010 STEM equipped with a a silicon drift detector from Bruker.

One of the samples was investigated by EBSD in order to obtain independent data about the structure of the upper Fe$_{3}$Si film.\cite{schwartz2009}  In the SEM Kikuchi-patterns are recorded point by point. The orientation of every point is recovered from the corresponding pattern with the limit of a lateral resolution of 20--30~nm. In this way the orientation distribution of the grains is determined in a very thin subsurface region.

\section{Results and Discussion}

\subsection{Al/Fe$_{3}$Si/GaAs multilayers}

\begin{figure}[!t]
\includegraphics[width=12.0cm]{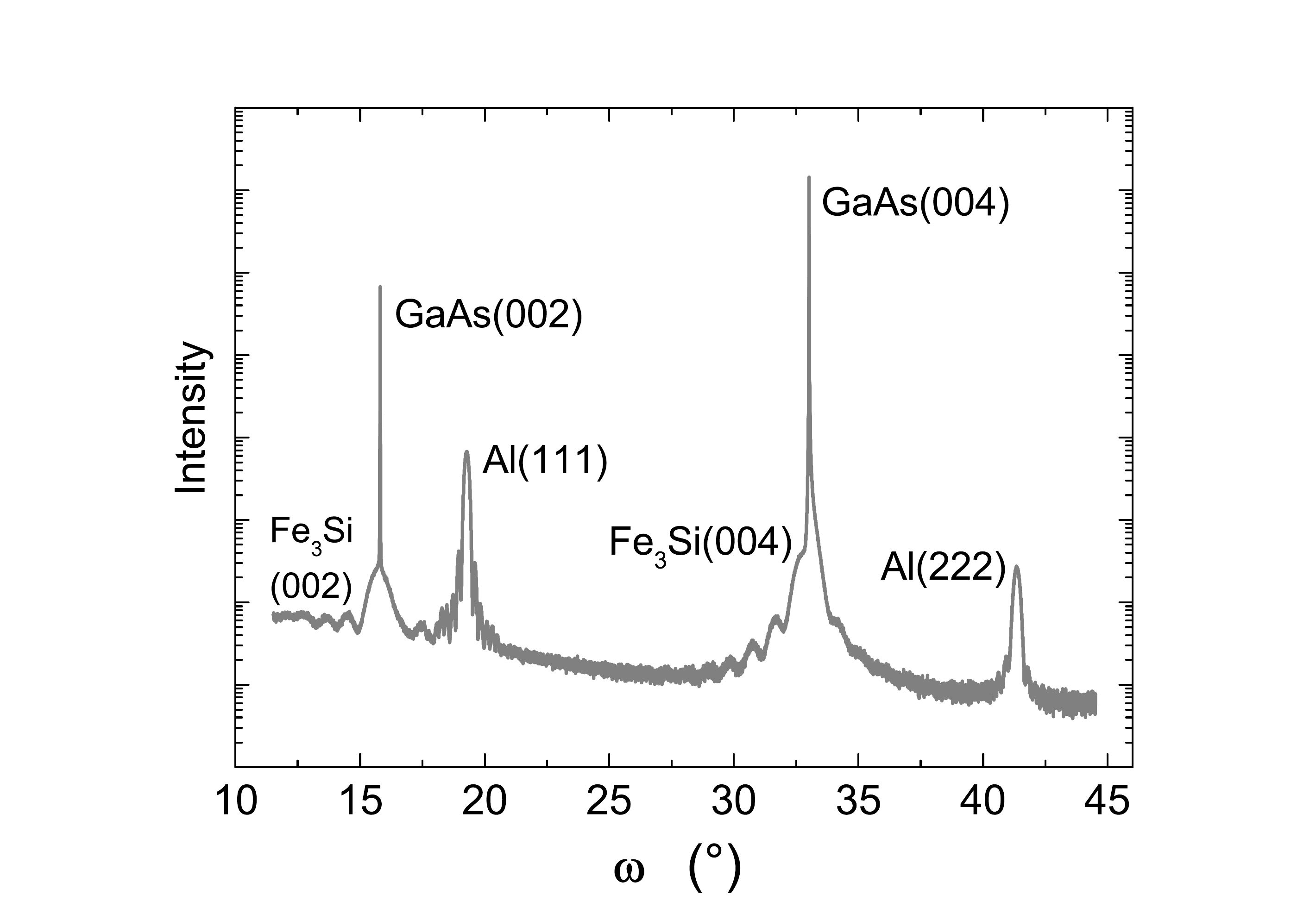}

\caption{HR~XRD diffraction curve of  sample 1 near the GaAs 002 and 004 reflections.}
\label{fig:XRD1_sample1}
\end{figure}

Figure~\ref{fig:XRD1_sample1} demonstrates the HR~XRD curve of sample~1 near the GaAs 002 and 004 reflections. The Fe$_{3}$Si 002 reflection is superimposed to GaAs 002 as the misfit between both lattices is very low, however, the layer peak is broader due to the finite film thickness. The corresponding 004 reflections behave in a similar manner. The strong peaks at higher angles correspond to the reflections Al 111 and 222.
The nominally 4.5-nm-thick Fe$_{3}$Si layer deposited at 200~$^\circ$C grows single-crystalline with a (001) cube-on-cube orientation as expected from our previous results.\cite{herfort03} The subsequent 23-nm-thick Al film deposited at 0~$^\circ$C grows almost \{111\}-oriented. If we only consider lattice matching, this fact is surprising since a coincidence lattice, achieved by a simple 45$^\circ$ in-plane rotation of the Al lattice (a~=~4.05~{$\AA$}) with respect to the Fe$_{3}$Si lattice (a~=~5.65~{$\AA$}) suggests a low (i.e. favorable) interface formation energy for the Al~\{001\} orientation. If we neglect the restrictions due to epitaxy, we can assume, that Al~\{111\} is energetically more favorable  thanks to the fact that the unstrained (111) plane has a 15~\% higher packing density than the (001)~plane. This implies a reduced energy of the (111) surface. Indeed, the ratio of surface energies of the fcc Al lattice for the 111 and the 100 directions has been calculated by density functional theory  to be 0.77.\cite{Vitos1998} We will return to this point later.

\begin{figure}[!t]
\includegraphics[width=12.0cm]{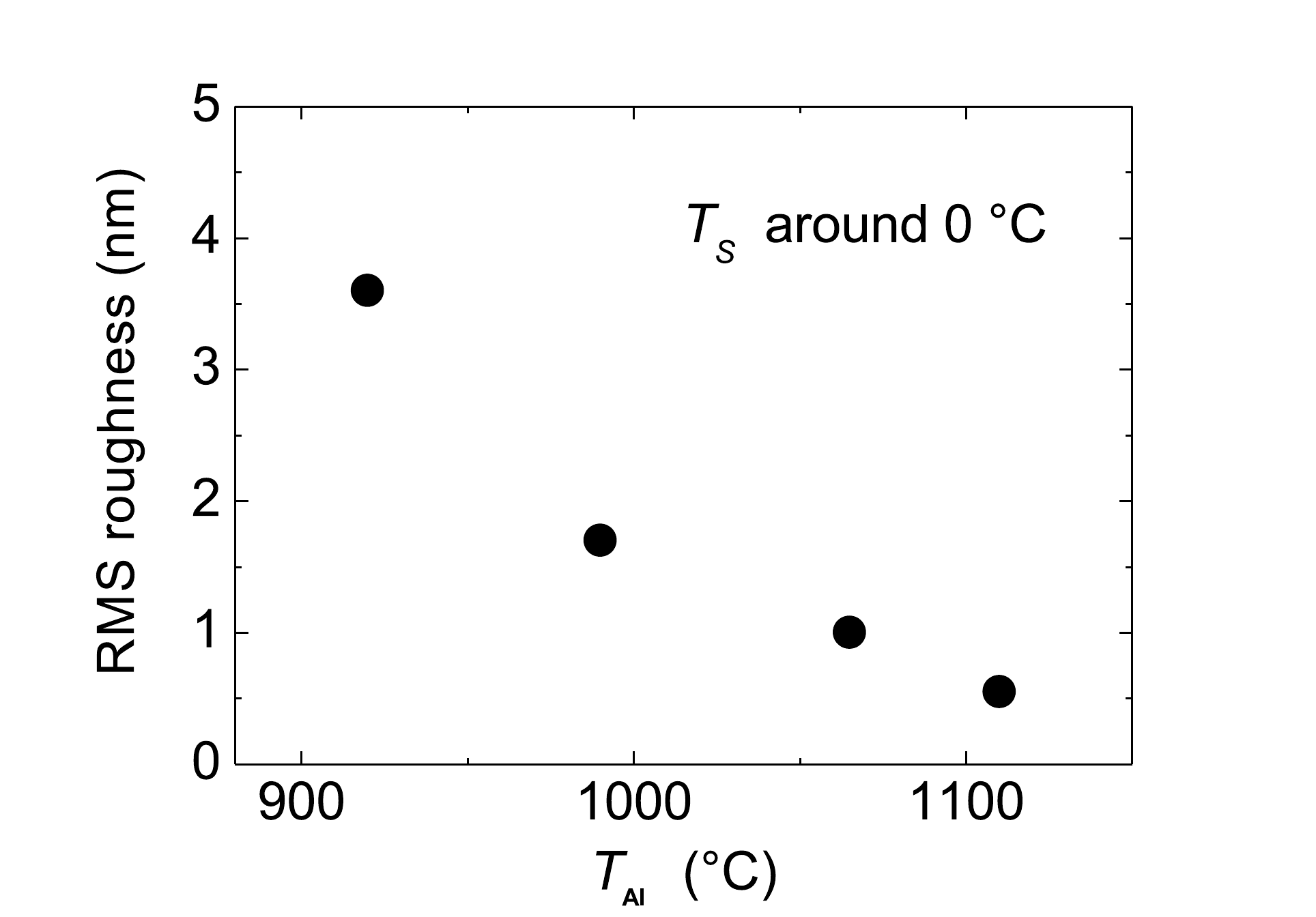}
\caption{RMS roughness of the Al surface measured by AFM over an area of 1~$\mu$m$^2$ in dependence of the Al source temperature, and the Al growth rate. }
\label{fig:RMS_rough_sample1}
\end{figure}

Figure~\ref{fig:RMS_rough_sample1}  demonstrates the RMS roughness of the Al surface in dependence of the Al-source temperature and hence the Al growth rate. The roughness is low over a large range of cell temperatures, higher cell temperatures yield lower RMS roughness. Hence, besides a low growth temperature of about 0~$^\circ$C a relatively high growth rate for the metals is required to obtain smooth Al films on Fe$_{3}$Si(001).

Analysis of the thickness fringes of the XRD peaks of sample~1 shown in Fig.~\ref{fig:XRD1_sample1} yields film thicknesses of (5 $\pm$ 0.5)~nm for Fe$_{3}$Si and (21 $\pm$ 1)~nm for Al very close to the nominal thicknesses. The slightly smaller Al thickness is due to the formation of a thin oxide layer of about 2--3~nm on top of the Al layer in accordance with x-ray reflectivity measurement (not shown here).

\begin{figure}[!t]
\includegraphics[width=12.0cm]{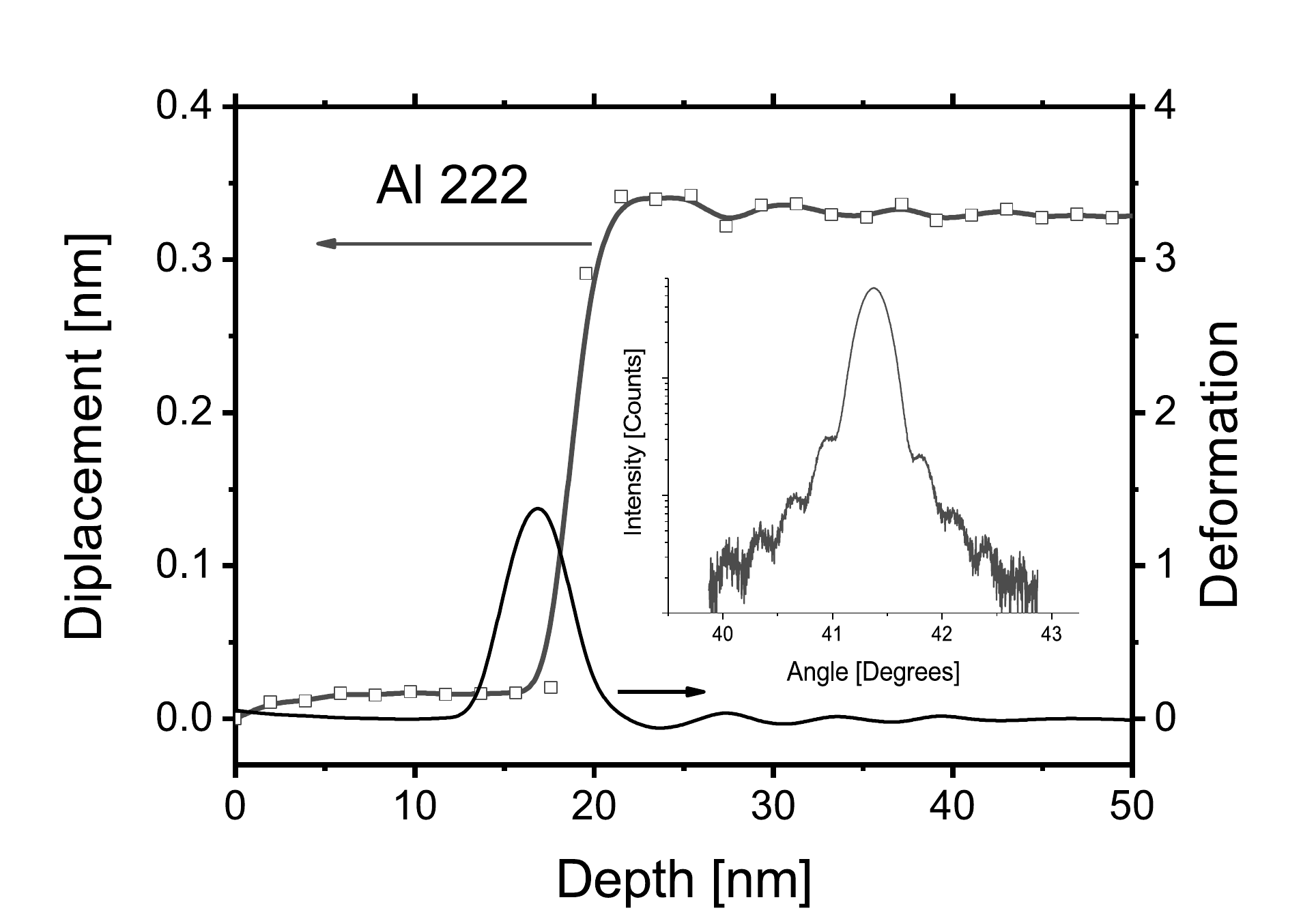}
\caption{Depth dependence of the displacement (line with open squares) and deformation (full line) inside the Al film obtained by phase retrieval of the Al 222 reflection. The HR~XRD diffraction curve is shown in the inset. The zero of the depth coordinate corresponds to the upper surface of the  Al film.}
\label{fig:XRD5}
\end{figure}

The 222 reflection of the Al film of sample~1 was analyzed by the x-ray phase retrieval method (see Fig.~\ref{fig:XRD5}) in order to obtain additional information about the Al growth on the Fe$_{3}$Si.~\cite{nikulin1998,nikulin1999} The 222 reflection of the Al film is shown in the inset. The result of the phase retrieval is the depth profile of the displacement (left axis). The zero of the depth coordinate corresponds to the upper surface of the  Al film. The phase retrieval was performed only for the Al-reflection and therefore the depth coordinate is connected only to the Al film. Sometimes it is more convenient to consider the depth profile of the deformation (right axis), which is the derivative of the displacement. As a result we obtain a (1.9 $\pm$ 0.5)~nm thick transition layer between Al and Fe$_{3}$Si at a depth of (18.4 $\pm$ 1)~nm.  We defined the transition layer as such: The displacement varies between 10$\%$ and 90$\%$ of the maximum value. We stress that the transition layer thickness was measured over an area of (1$\times$10)~mm$^2$. Later we will see that this result corresponds to more local measurements. The application of the phase retrieval method is justified because the coherence length of the radiation is in the $\mu$m range thanks to application of crystal optics elements.\cite{kaganer2001} Such a transition layer seems to justify the assumption, that the Al growth is not purely epitaxial and the misfit is accommodated in an nearly amorphous region. Similar transition layers have been observed for SrTiO$_3$ epitaxial films grown on Si(100).\cite{hu2003}

\begin{figure}[!t]
\includegraphics[width=12.0cm]{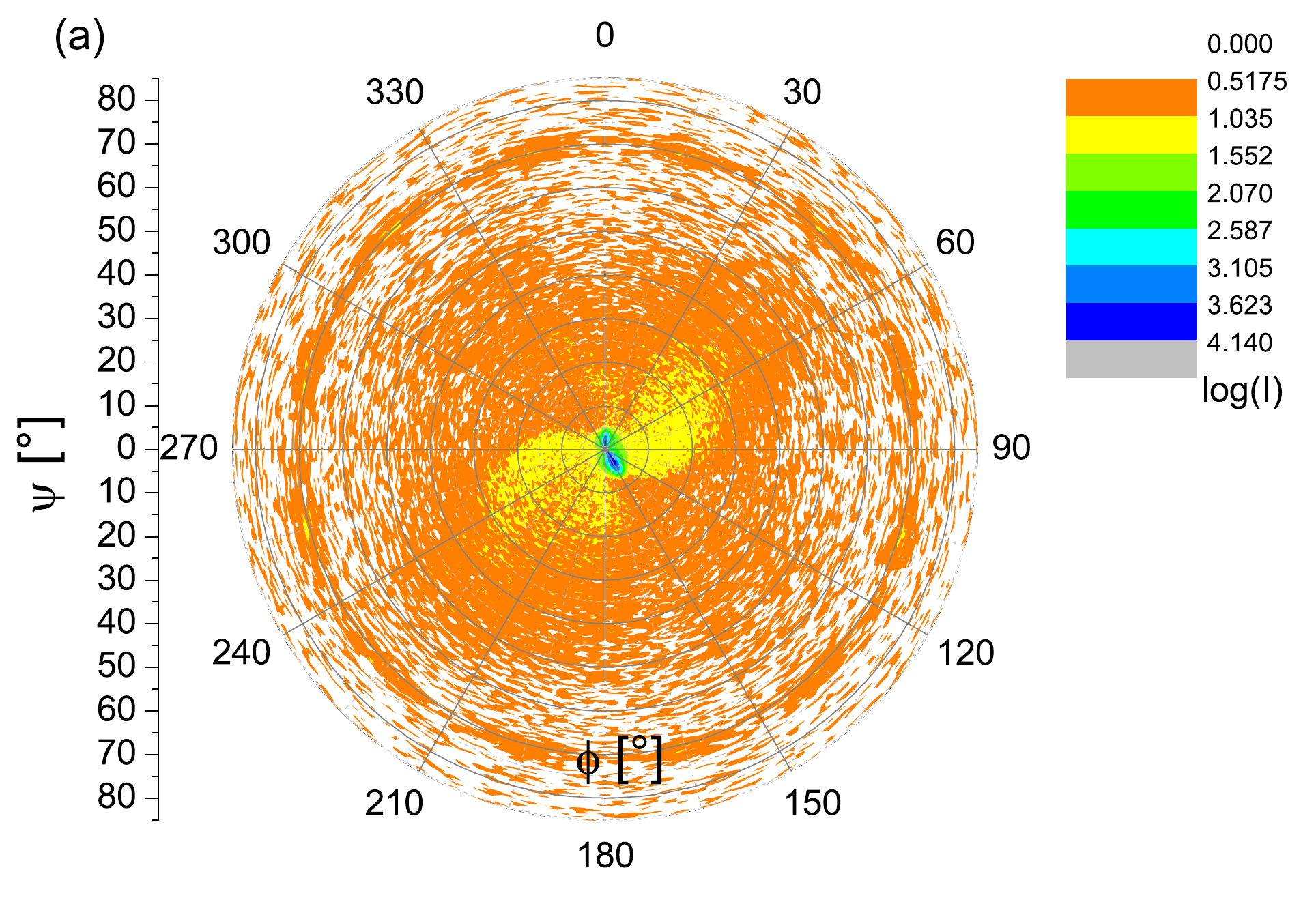}
\includegraphics[width=12.0cm]{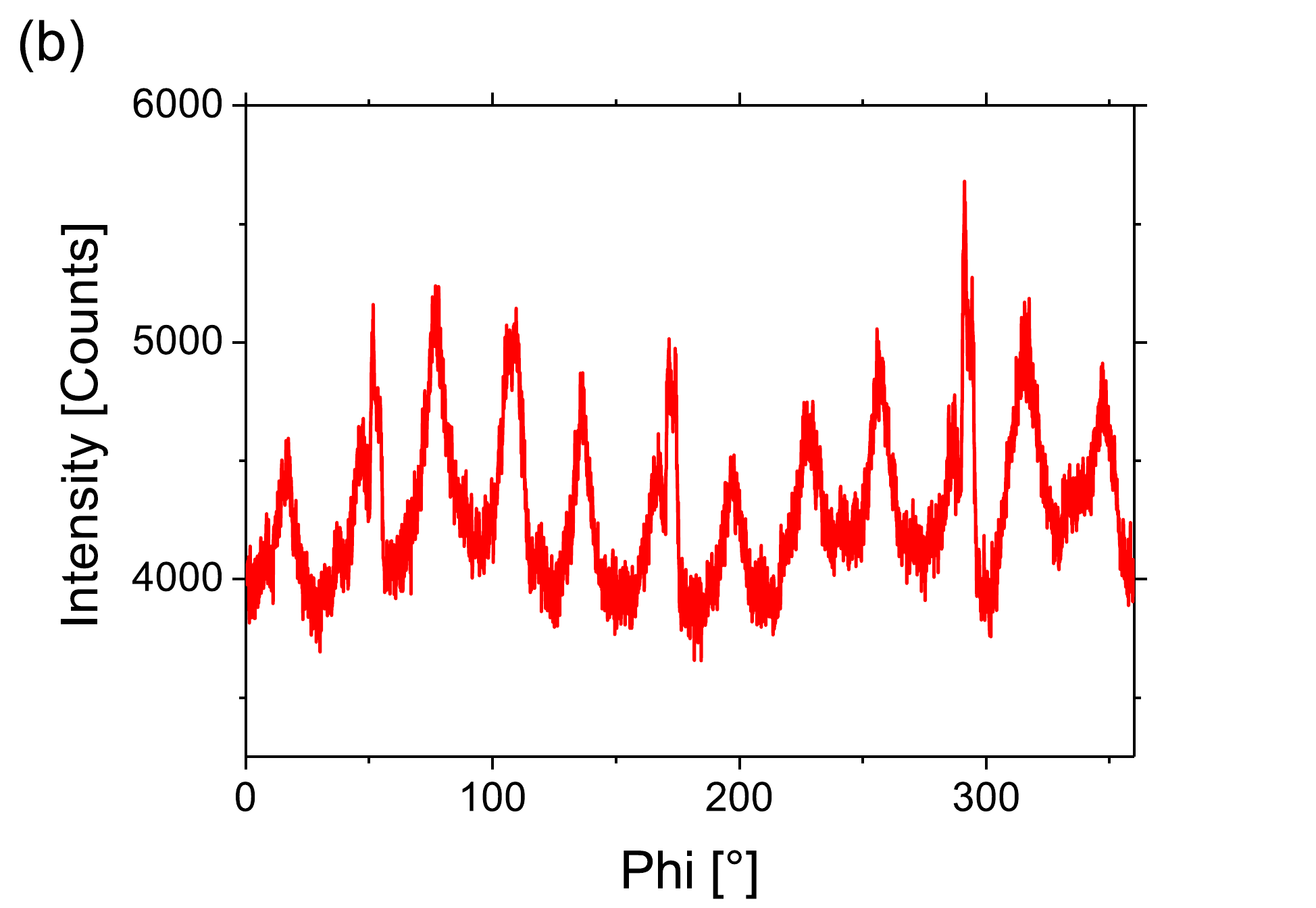}
\caption{(a) X-ray orientation map of the Al 111 reflection of sample~1. A strong fibre texture is found. The diffracted intensity is plotted here on a logarithmic scale. $\phi$ is the azimuthal angle of the sample and $\psi$ is the tilt angle. Near $\psi$ = 70$^\circ$ a ring with a somewhat enhanced intensity is observed. (b) Skew X-ray  measurement ($\phi$-scan) of the Al 111 reflection tilted by 70.5$^\circ$ with respect to the surface normal.}
\label{fig:XRD2}
\end{figure}

Figure~\ref{fig:XRD2} (a) shows the result of an XRD texture measurement of the Al 111 reflection (sample~1). A fibre texture is found. All the diffracted intensity remains in an angular range well below 10 degrees (first ring). The diffracted intensity is plotted here on a logarithmic scale, where $\phi$ is the azimuthal angle of the sample and $\psi$ is the tilt angle. Near $\psi$ = 70.5$^\circ$ a ring with a somewhat enhanced intensity is observed.
Therefore Fig.~\ref{fig:XRD2} (b) shows a $\phi$-scan of the Al 111 reflection tilted by 70.5$^\circ$ (sample~1). 12 maxima are observed, whereas only 3 maxima are expected for a single crystal  (regarding to the threefold symmetry of the 111 lattice). Consequently the Al film consists of four different structural domains leading to 12~peaks found in the experiment.

\begin{figure}[!t]
\includegraphics[width=8.0cm]{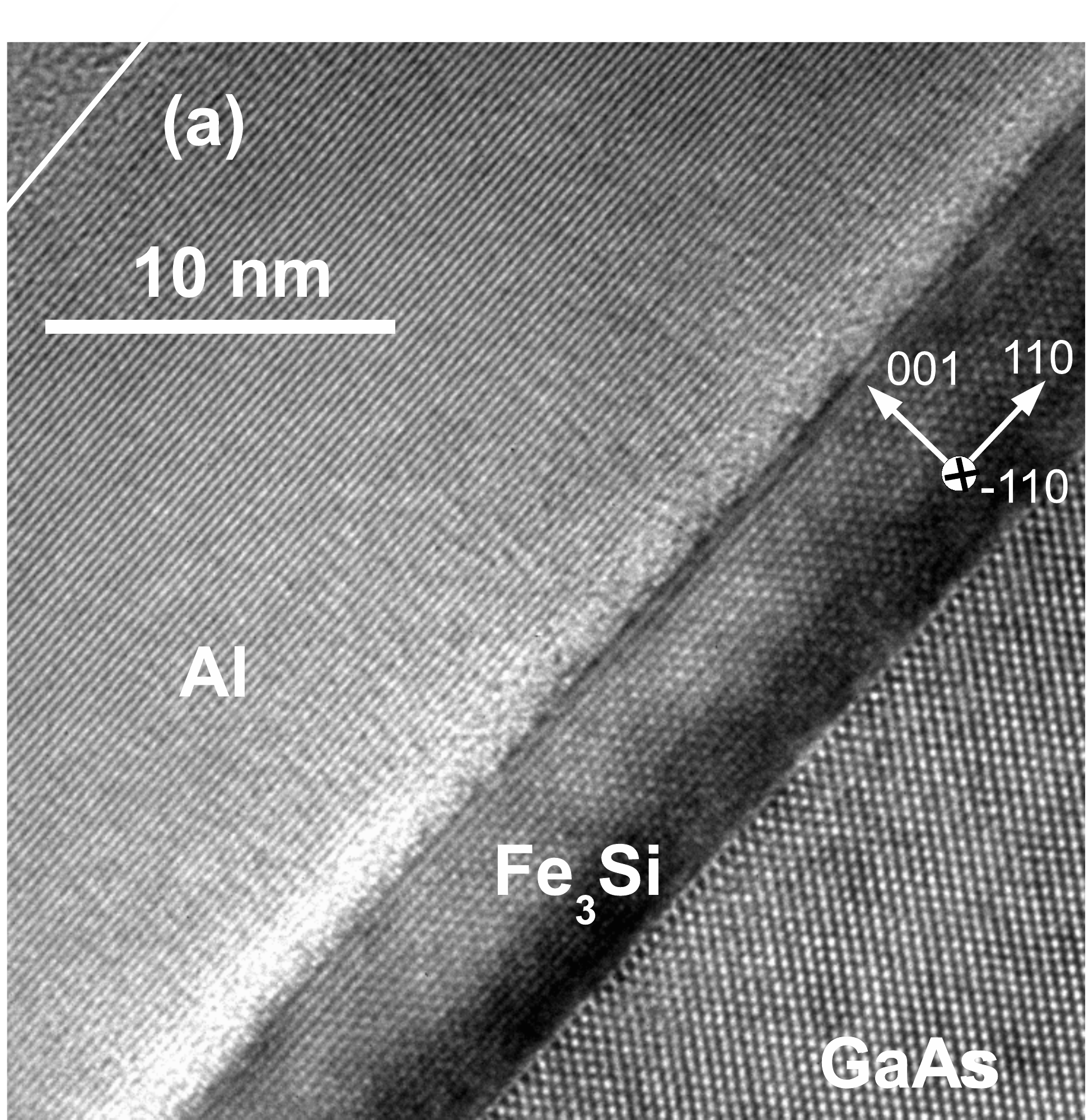}
\includegraphics[width=8.0cm]{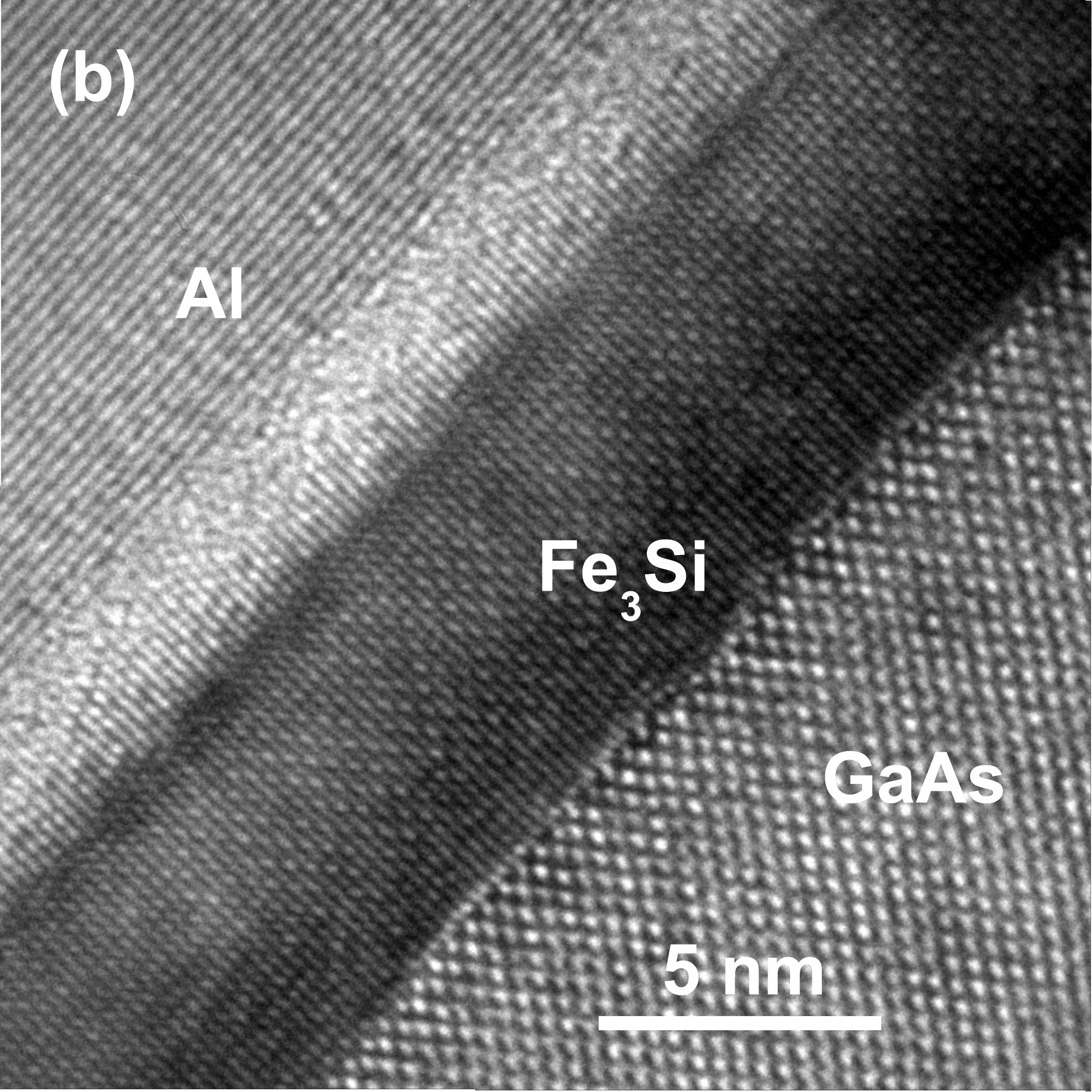}
\includegraphics[width=8.0cm]{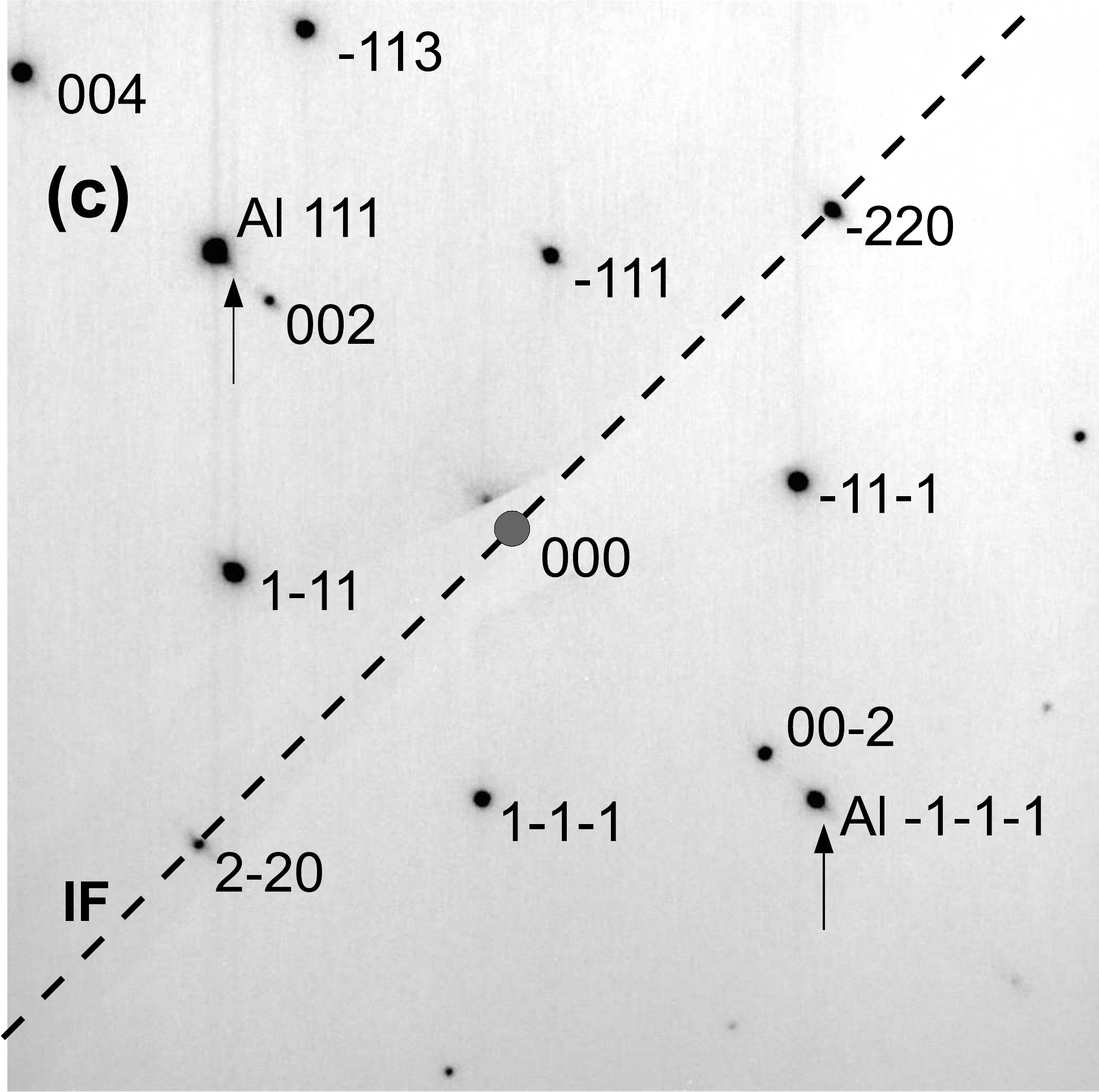}
\caption{(a) Cross-section HR~TEM micrograph of sample 1.  The Fe$_{3}$Si-film is an epitaxial film. The white line marks the upper surface of the Al film. (b) Higher magnification micrograph demonstrating the high quality structure of the Fe$_{3}$Si and the IFs in more detail. (c) Corresponding selected area diffraction pattern, exhibiting the spots due to GaAs and Fe$_{3}$Si (marked) together with only two spots due to Al 111 and Al $\bar{1}$$\bar{1}$$\bar{1}$. The vertical lines are artefacts.}
\label{fig:TEM1}
\end{figure}

\begin{figure}[!t]
\includegraphics[width=12.0cm]{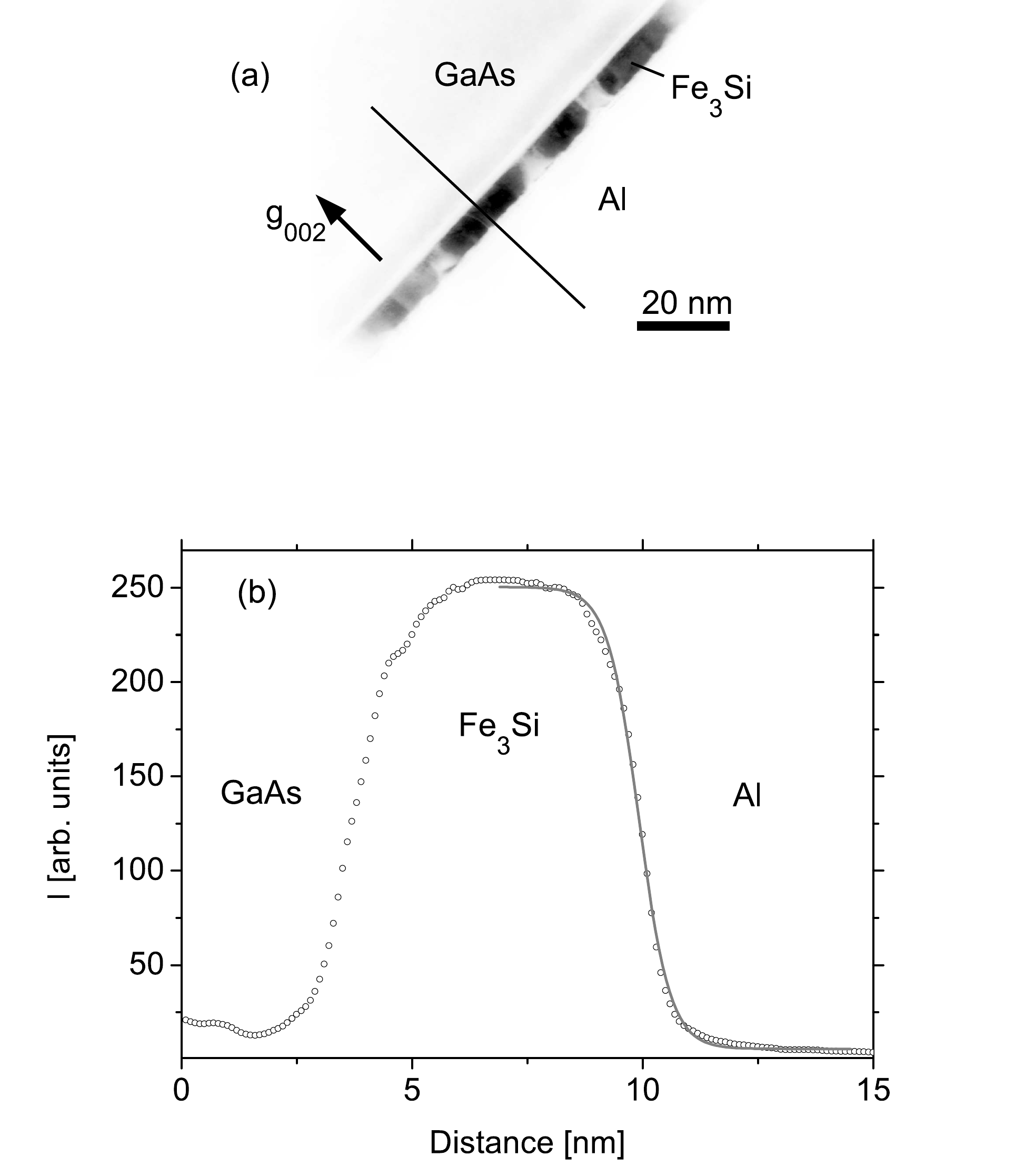}
\caption{(a) DF TEM micrograph of sample 1 using the Fe$_{3}$Si 002 reflection (inverted image). The location of the line profile of the diffracted intensity given in (b) is marked. The experimental line profile (open circles) was fitted by a sigmoidal function (full line). The zero coordinate of the line profile is chosen arbitrarily.}
\label{fig:TEM0}
\end{figure}

Figure~\ref{fig:TEM1}~(a) shows a cross-section HR~TEM micrograph of sample 1. The GaAs/Fe$_{3}$Si interface (IF) is smooth, whereas the Fe$_{3}$Si/Al IF appears slightly more rough. The Fe$_{3}$Si-film is  epitaxial with lateral grain-like inhomogeneities on a length-scale of about 10~nm [see Fig.~\ref{fig:TEM0}~(a)]. The 111 planes of the Al-film are well oriented parallel to the IF, although the in-plane orientation of the Al-film seems to be arbitrary, because only interference fringes parallel to the IF are found. The Fe$_{3}$Si film thickness is (5.9 $\pm$ 0.5)~nm and the Al film is (20 $\pm$ 1)~nm thick. The white line marks the upper surface of the Al film. Figure~\ref{fig:TEM1}~(b) depicts a higher magnification micrograph demonstrating the high structural quality  of the Fe$_{3}$Si and both IFs. Inside the imaged lateral region the transition layer between Fe$_{3}$Si and Al can be distinguished to be (1.5 $\pm$ 0.5)~nm thick. Within this transition layer the interference fringes due to net-planes perpendicular to the IF vanish gradually, probably due an increase of disorder in the Fe$_{3}$Si film. Above the transition layer the interference fringes due to netplanes parallel to the IF arise. Figure~\ref{fig:TEM1}~(c) shows the corresponding selected area diffraction (SAD) pattern, exhibiting the spots due to GaAs and Fe$_{3}$Si together with only two additional spots due to Al 111 and Al $\bar{1}$$\bar{1}$$\bar{1}$. Streaks perpendicular to the IF (marked by arrows) are observed in the vicinity of the strongest maxima indicating a smooth surface and IFs. From the analysis of the dark-field (DF) TEM micrograph shown in Fig.~\ref{fig:TEM0} the transition layer between Al and Fe$_{3}$Si was characterized using fits of the intensity line-profile by a sigmoidal function \cite{luna2009}. The transition layer is (1.8 $\pm$ 0.5)~nm thick. According to our definition of the transition layer the diffracted intensity varies between 10$\%$ and 90$\%$ of the maximum value. These results correspond well to those obtained by x-ray phase retrieval, described earlier.

\begin{figure}[!t]
\includegraphics[width=12.0cm]{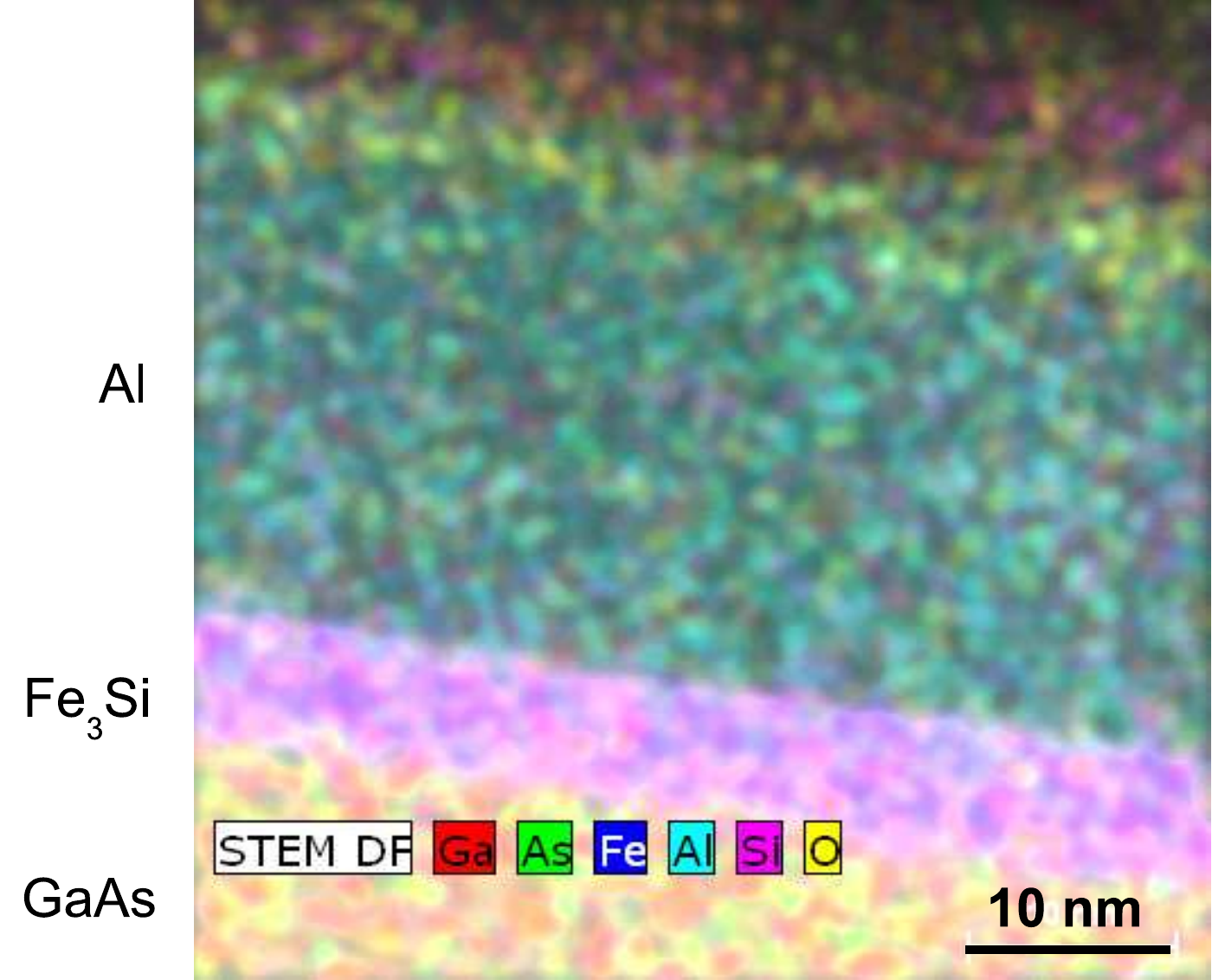}
\caption{EDX elemental map of sample 1 obtained in the STEM. The Fe$_{3}$Si/Al interface is chemically more abrupt than the Fe$_{3}$Si/GaAs interface.}
\label{fig:EDX}
\end{figure}

In addition we performed EDX measurements in the STEM in order to study the composition of the films.
Figure~\ref{fig:EDX} shows an EDX elemental map of sample~1. The Fe$_{3}$Si/Al interface is chemically more abrupt than the Fe$_{3}$Si/GaAs interface. This fact may be explained by the difference in deposition temperature. The Al-film was deposited at the low substrate temperature T$_s$~=~0~$^\circ$C whereas the Fe$_{3}$Si film was  grown at T$_s$~=~200~$^\circ$C. The higher temperature leads to enhanced diffusion. From the EDX measurement of the Al distribution perpendicular to the interfaces we obtained an IF thickness of (1.5 $\pm$ 0.5)~nm. The Al concentration in the transition layer is varying linearly. This clearly shows that the conception of the interface layer is restricted mainly to structural properties and no anomaly of the Al concentration profile is found.

The formation of the transition layer, which elastically accommodates the strain, decouples the Al layer from the registry of the underlying Fe$_{3}$Si layer and triggers the growth of the Al in (111) orientation, which is now energetically more favorable as mentioned earlier. Note, that we did not find misfit dislocations or other defects connected to strain relaxation. Similar phenomena were applied using oxide hetero-structures as buffers for the integration of alternative semiconductors.\cite{schroeder2009}

\subsection{Fe$_{3}$Si/Al/Fe$_{3}$Si/GaAs multilayers}

\begin{figure}[!t]
\includegraphics[width=12.0cm]{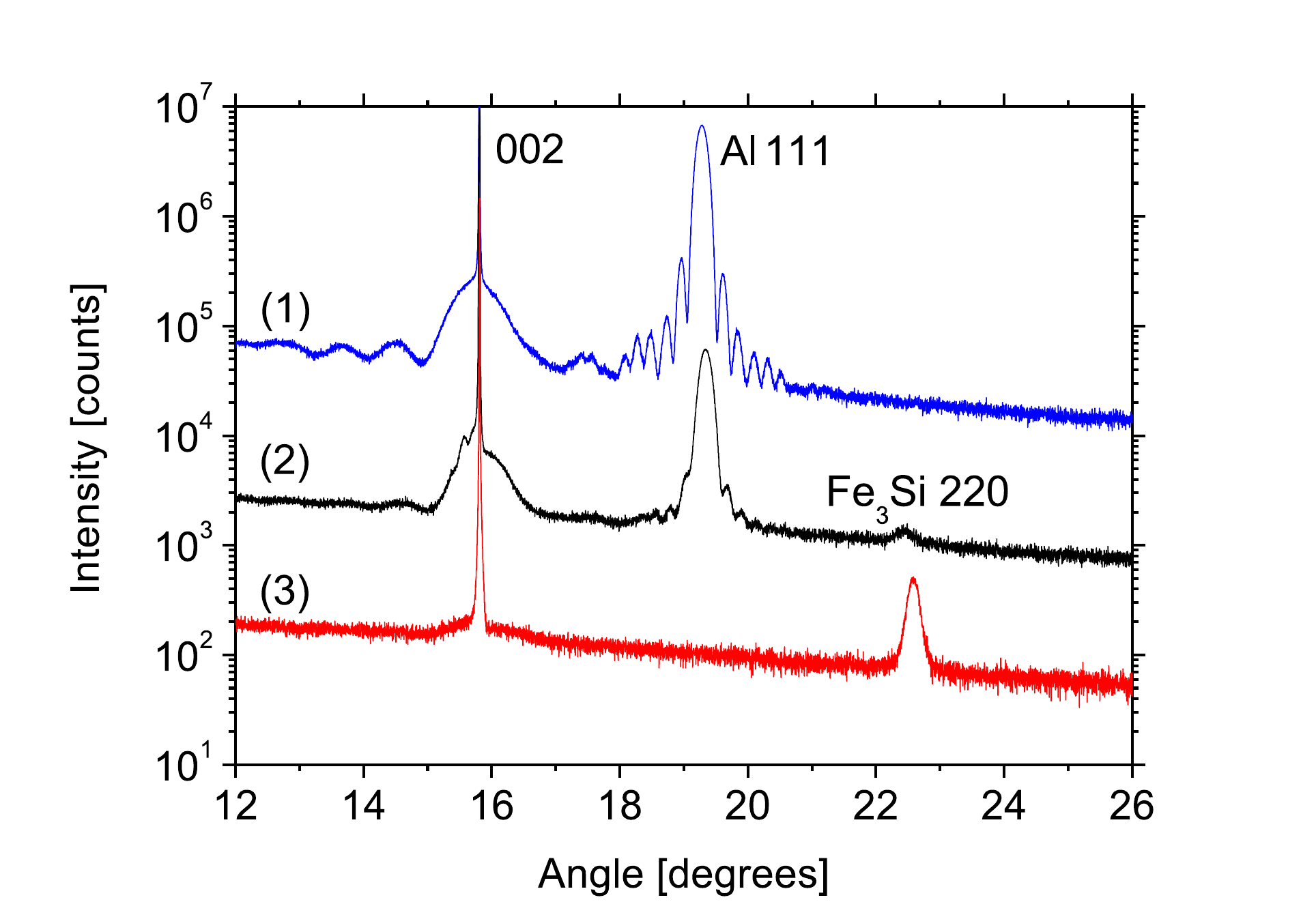}
\caption{ HR~XRD diffraction curves of all the samples near the GaAs 002 reflection. The Al peak near 19.2$^\circ$ is missing for sample~3 and the Fe$_{3}$Si 220 reflection is observed for the samples (2) and (3).}
\label{fig:XRD1}
\end{figure}

Figure~\ref{fig:XRD1} demonstrates the HR~XRD curves of all the samples near the GaAs 002 reflection. Superimposed  Fe$_{3}$Si 002 reflection and GaAs 002 reflection occur. The other strong peak corresponds to Al 111.
The lower 4.5~nm-thick Fe$_{3}$Si layer deposited at 200$^\circ$C grows single-crystalline with a \{001\} orientation. The subsequent 23~nm-thick Al film deposited at 0$^\circ$C grows almost \{111\}-oriented. The peak at 22.57$^\circ$ corresponds to  Fe$_{3}$Si 220 originating from the upper Fe$_{3}$Si film.

\begin{figure}[!t]
\includegraphics[width=8.0cm]{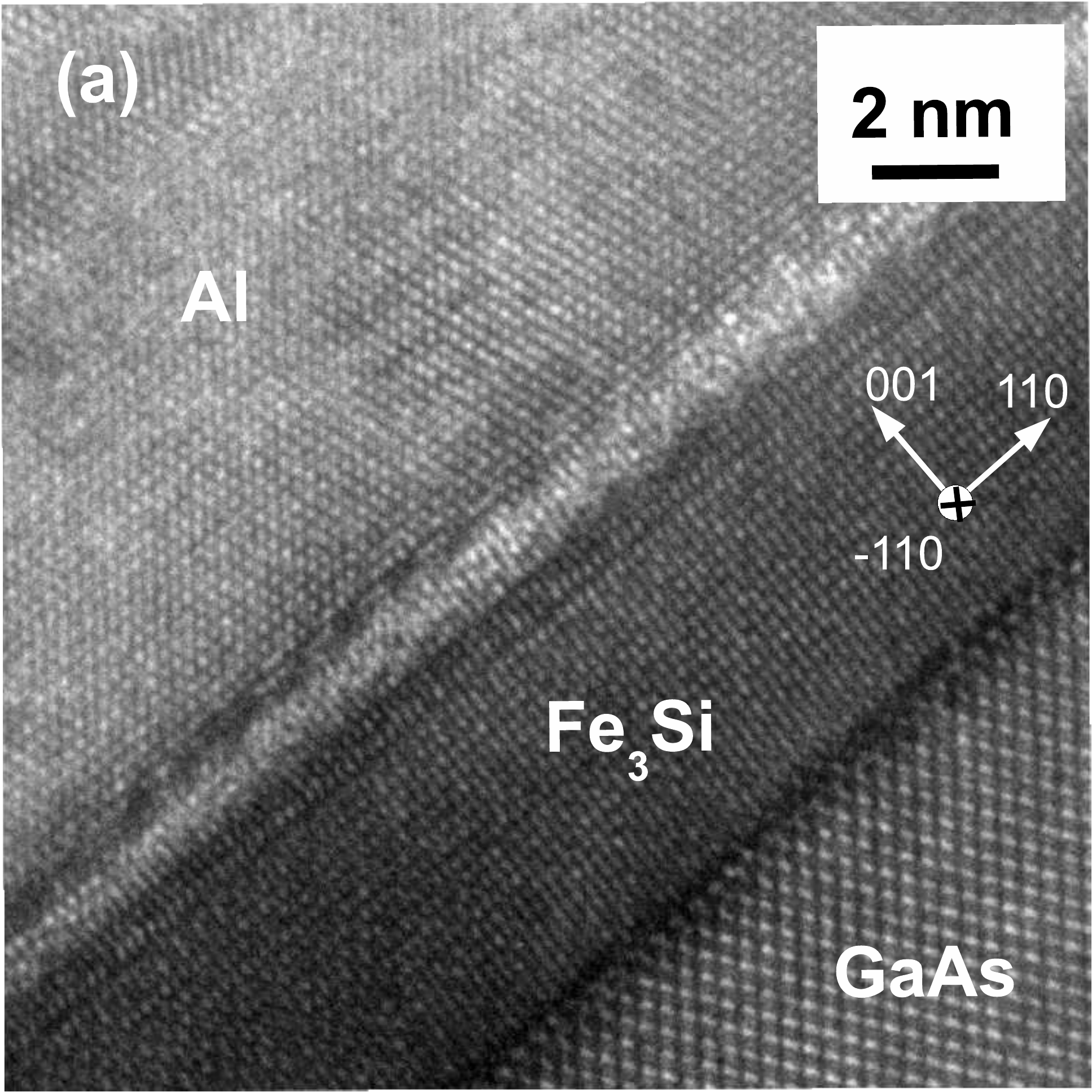}
\includegraphics[width=8.0cm]{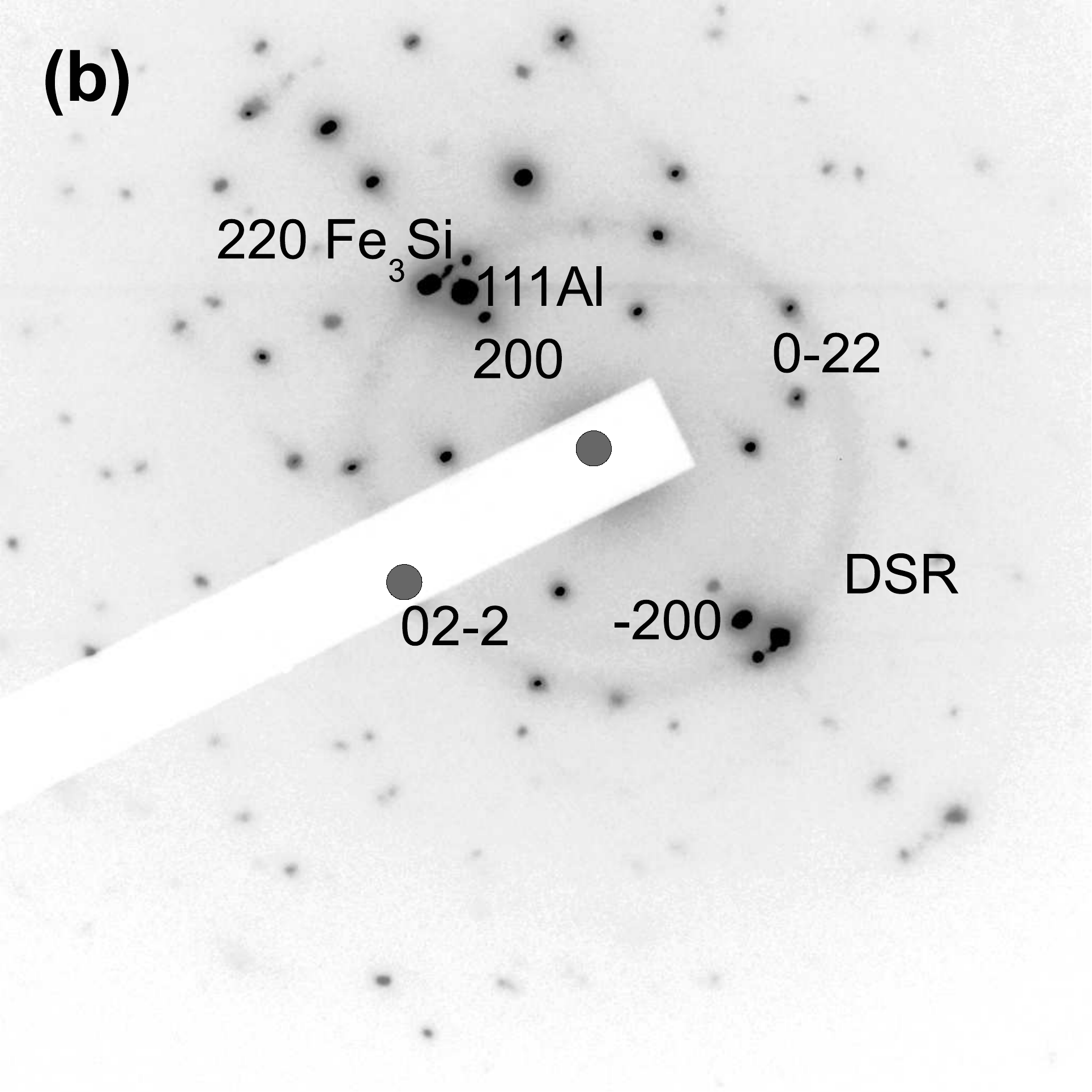}
\includegraphics[width=8.0cm]{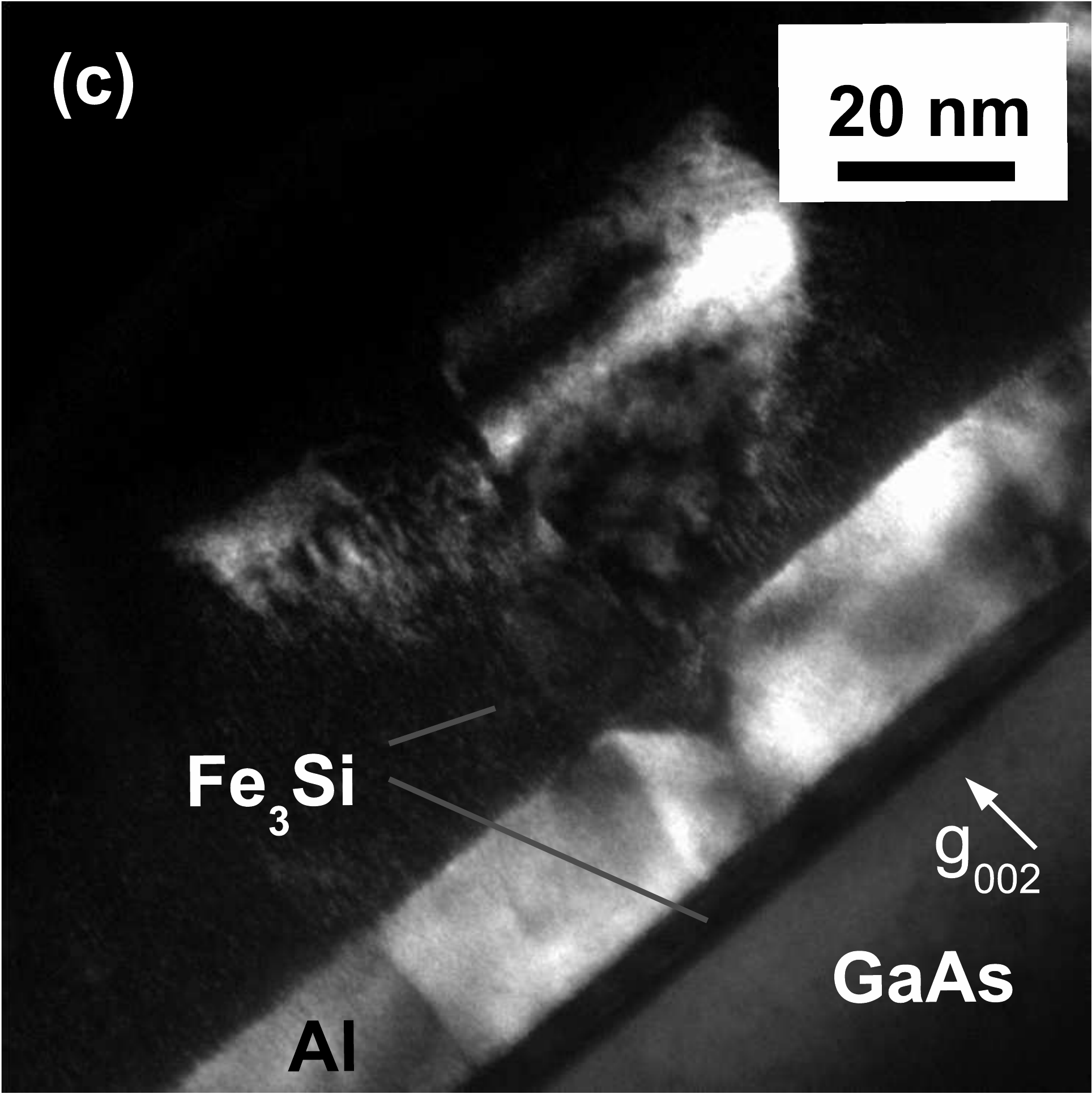}
\caption{(a) Cross-section HR~TEM micrograph of sample~2. The first Fe$_{3}$Si-film is an epitaxial film. For the Al film we observe locally a perfect interference pattern, which is tilted by $\simeq$10$^\circ$ with respect to the substrate pattern around the axis [$\overline{1}$10]. (b) Corresponding selected area diffraction pattern of sample~2. (c) Corresponding DF micrograph at lower magnification of sample~2. The Al film and the upper Fe$_{3}$Si film are textured poly-crystals}
\label{fig:TEM2}
\end{figure}

\begin{figure}[!t]
\includegraphics[width=12.0cm]{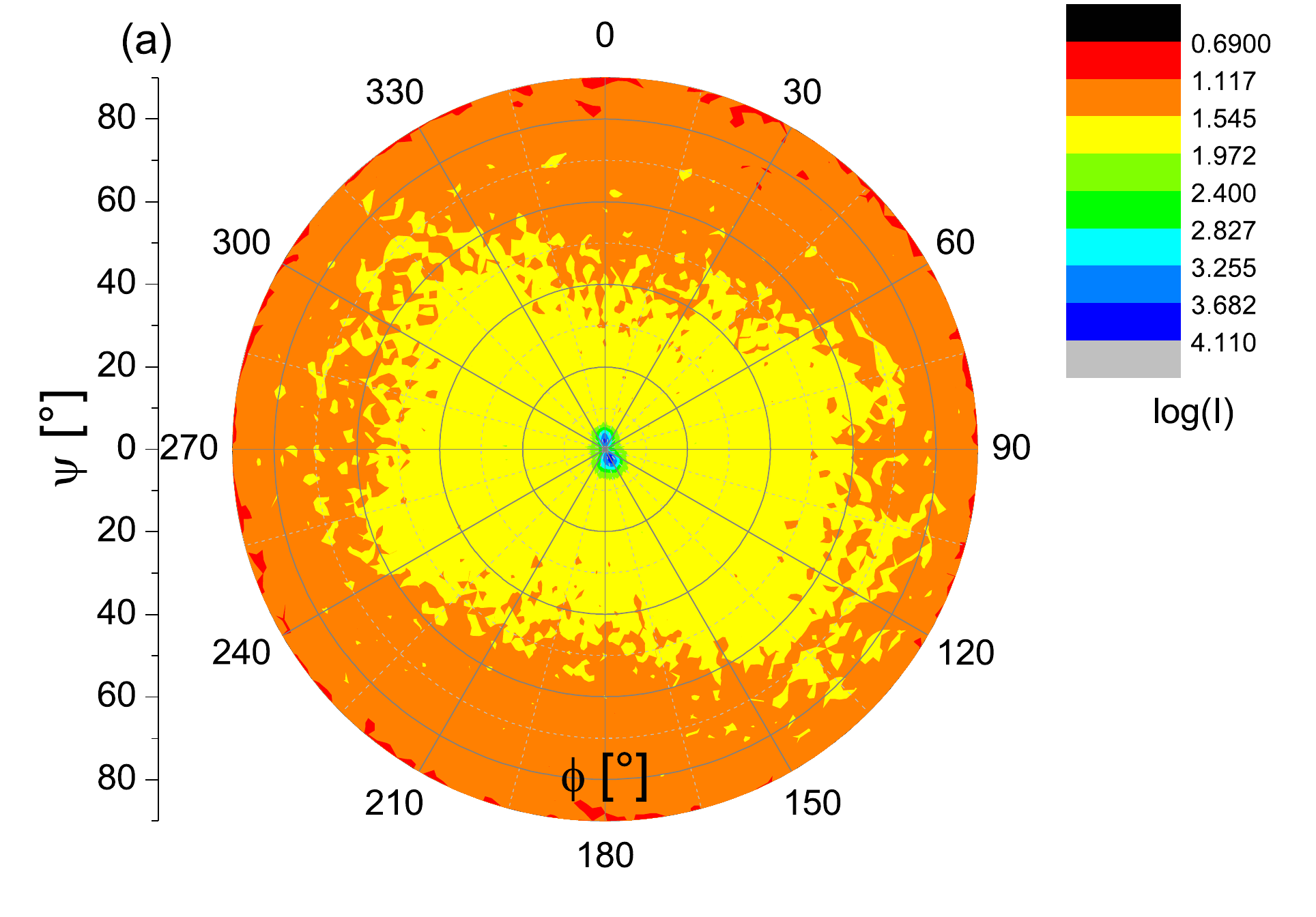}
\caption{a) X-ray orientation map of the Al 111 reflection of sample~2. A fibre texture is found. The diffracted intensity is plotted here on a logarithmic scale. $\phi$ is the azimuthal angle of the sample and $\psi$ is the tilt angle. }
\label{fig:XRD4}
\end{figure}

Figure~\ref{fig:TEM2}~(a) shows a cross-section HR~TEM micrograph of sample~2. The IFs are smooth both below and above the lower Fe$_{3}$Si film, i.e. the GaAs/Fe$_{3}$Si IF as well as the Fe$_{3}$Si/Al IF.  The first Fe$_{3}$Si-film is epitaxially aligned similar as that of sample~1. For the Al film we observe locally a perfect interference pattern. Here a (1.1 $\pm$ 0.5)~nm thick transition layer is detected at the Al/Fe$_{3}$Si  IF similar to the previously mentioned results. Figure~\ref{fig:TEM2}(b) shows a corresponding SAD pattern of sample~2. Besides the GaAs 200 peak the Al 111 and the Fe$_{3}$Si 220 reflections are found in close vicinity, i.e. the orientation of the upper Fe$_{3}$Si film is completely different compared to the lower one. The splitting of the Fe$_{3}$Si 220 peak and a Debye-Scherrer-Ring (DSR) are observed, i.e. the upper Fe$_{3}$Si is not an epitaxial film. Figure~\ref{fig:TEM2} (c) shows the corresponding DF micrograph at lower magnification of sample~2 depicting the whole layer stack. Both upper films (Al and Fe$_{3}$Si) show a grainy structure.
Figure~\ref{fig:XRD4}~(a) exhibits an X-ray orientation map of the Al 111 reflection of sample~2. A fibre texture has been found. The diffracted intensity is plotted here on a logarithmic scale.
The overgrowth of the Al film with Fe$_{3}$Si at low substrate temperature does not harm the overall quality of the multi-layer structure. In order to improve the structural properties of the uppermost Fe$_{3}$Si layer it might be desirable to increase the Fe$_{3}$Si growth temperature towards 200~$^\circ$C (the optimum T$_S$ for Fe$_{3}$Si on GaAs). However, as we will show in the following an increase of T$_S$ leads to a strong degradation of the whole stack.
The system Fe$_{3}$Si/GaAs is stable during annealing up to 425~$^\circ$C. So, a suitable post-growth annealing step could possibly further improve the crystallinity of the layer stack. During MBE growth of Fe$_{3}$Si the system Fe$_{3}$Si/GaAs(001) is stable only up to a temperature of 250~$^\circ$C,\cite{herfort03} i.e. MBE deposition and high temperature annealing have a different influence on structural quality of the layer stack.

\begin{figure}[!t]
\includegraphics[width=12.0cm]{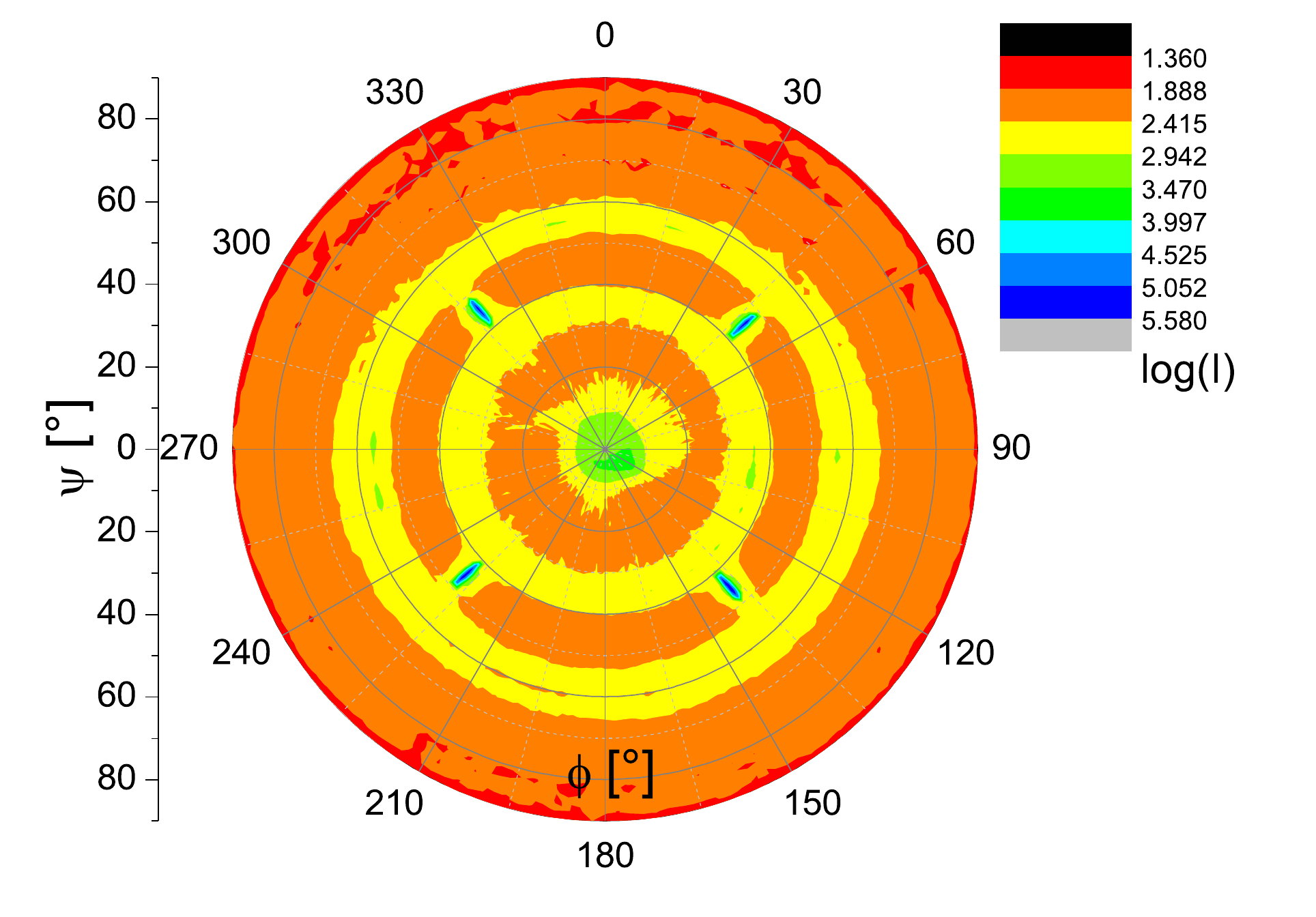}
\caption{(a) X-ray orientation map of the Fe$_{3}$Si 220 reflection of sample~3. A fibre texture is found in the second Fe$_{3}$Si film. The diffracted intensity is plotted here on a logarithmic scale. $\phi$ is the azimuthal angle of the sample and $\psi$ is the tilt angle. Near $\psi$ = 45$^\circ$ peaks from the substrate and the first Fe$_{3}$Si film are observed.}
\label{fig:XRD6}
\end{figure}

\begin{figure}[!t]
\includegraphics[width=8.0cm]{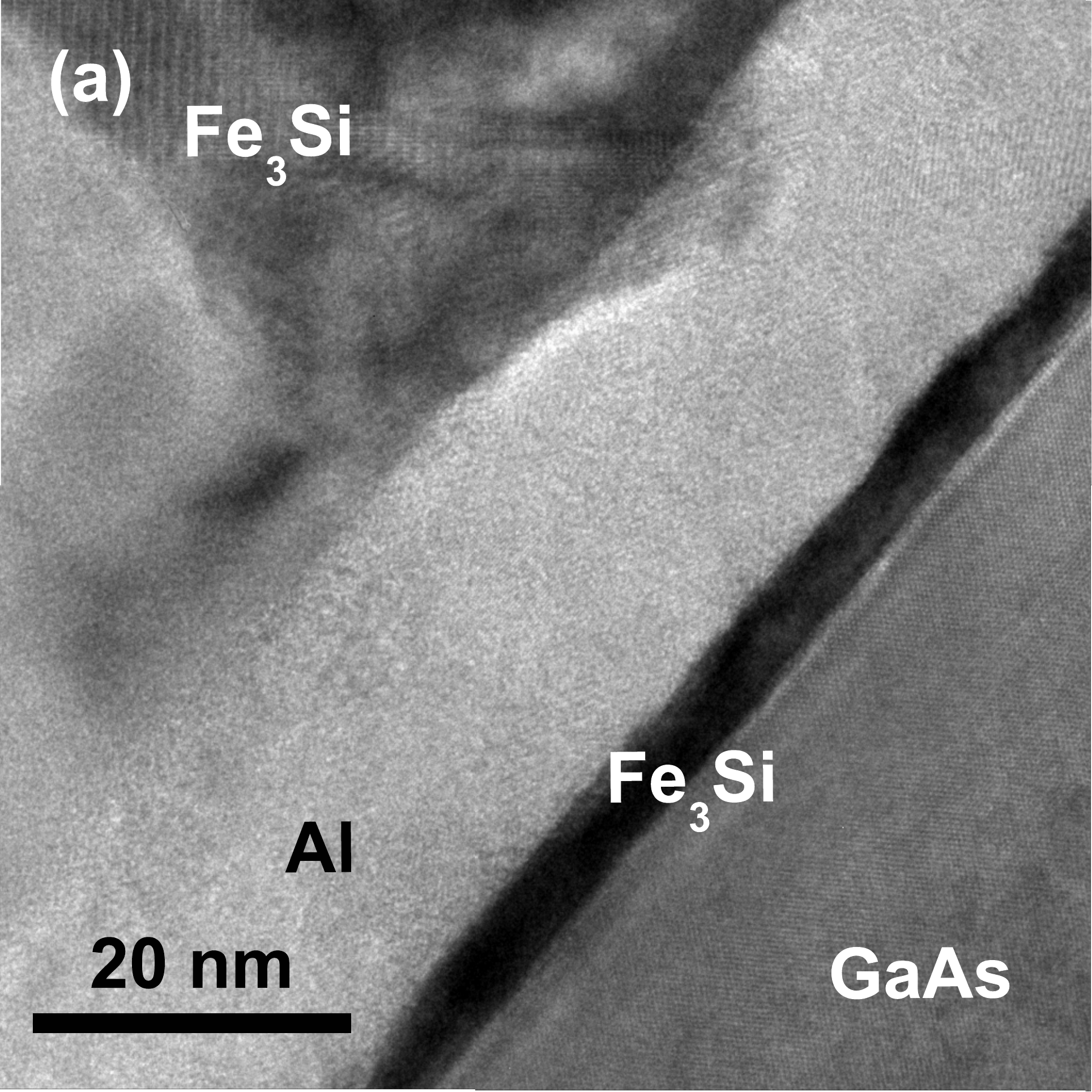}
\includegraphics[width=8.0cm]{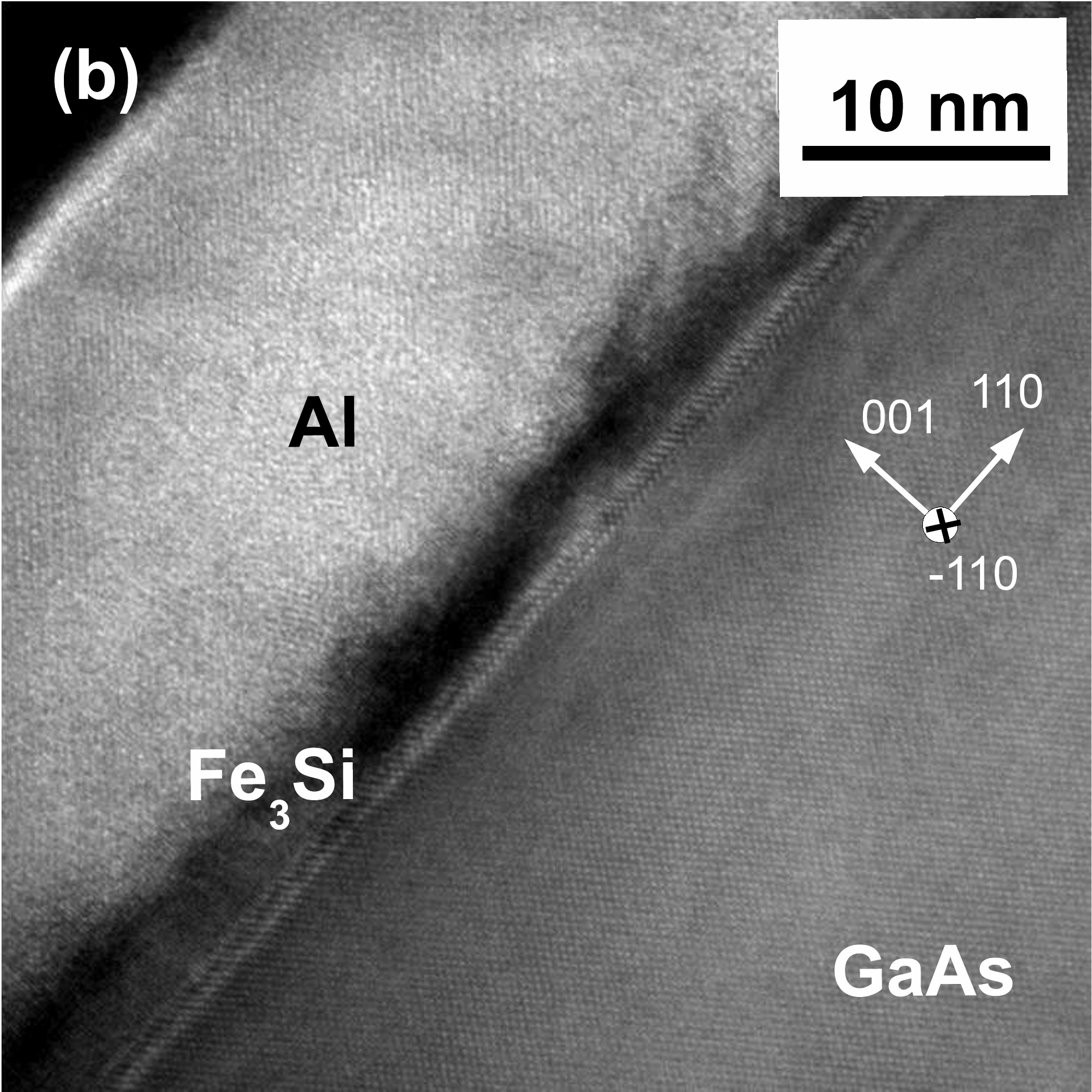}
\includegraphics[width=8.0cm]{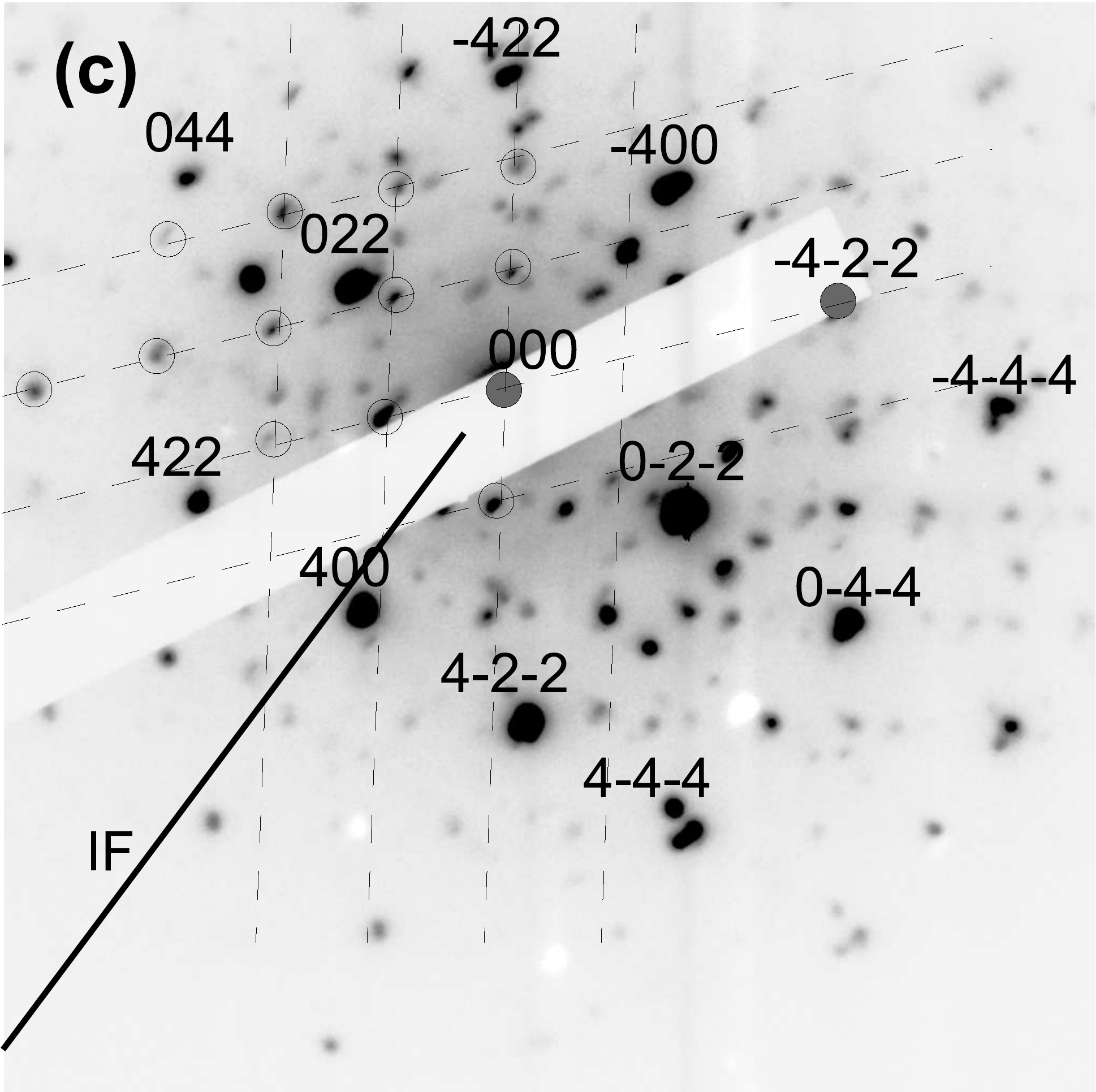}
\caption{(a) Cross-section HR~TEM micrograph of sample~3. The Al film and the upper Fe$_{3}$Si layer are polycrystalline although some lattice planes still seem to be visible. (b) Cross-section HR~TEM micrograph of sample~3 at higher magnification. (c) SAD pattern of sample~3, Fe$_{3}$Si maxima are indexed, substrate maxima are indicated by dashed lines and open circles. The direction of the IF is given by the full line.}
\label{fig:TEM3}
\end{figure}

\begin{figure}[!t]
\includegraphics[width=10.0cm]{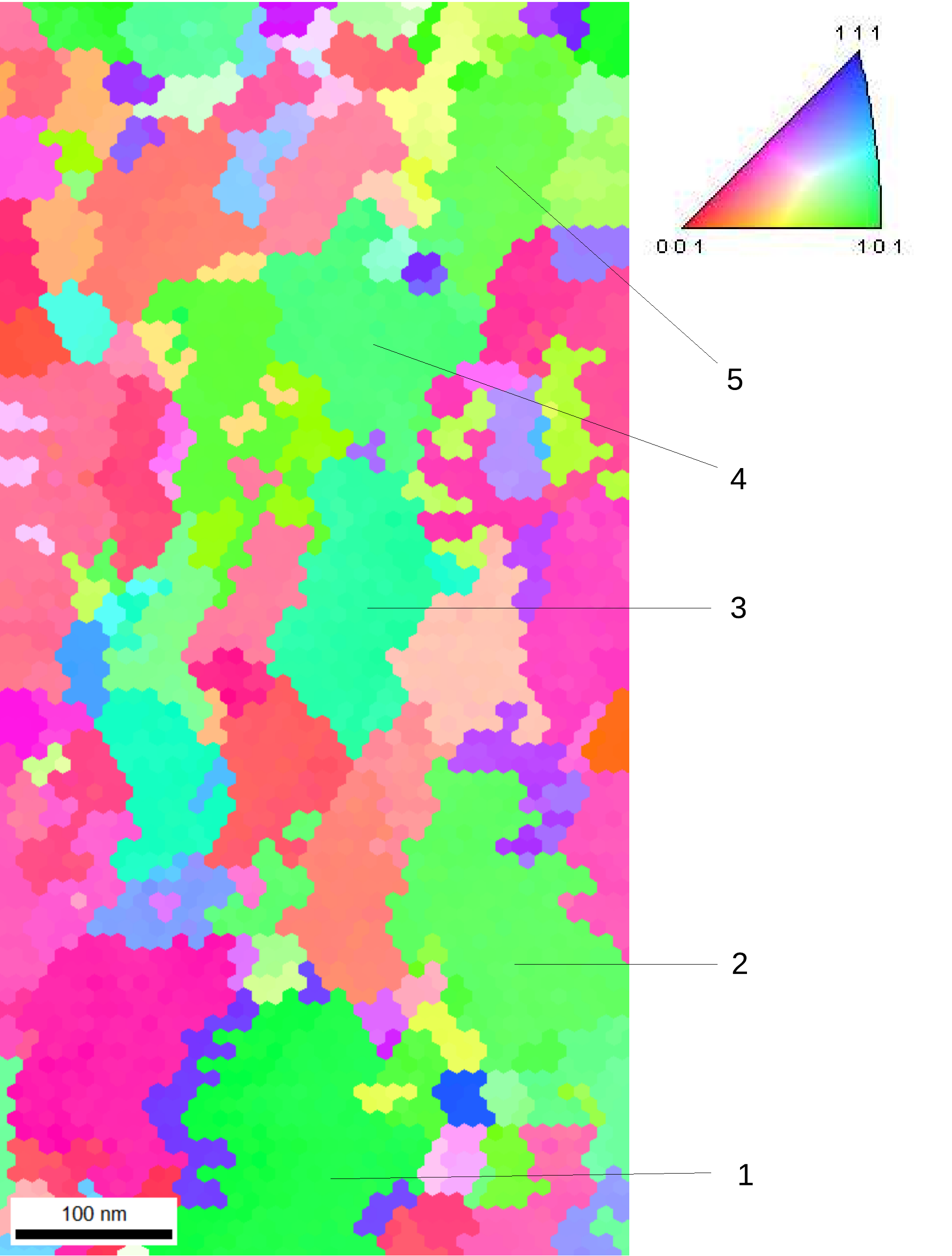}
\caption{SEM EBSD micrograph of sample~3 illustrating the orientation of the different grains of the upper Fe$_{3}$Si film. The numbered grains have a {101} orientation perpendicular to the substrate surface.}
\label{fig:ebsd}
\end{figure}

The XRD curve of sample~3 shown in Fig.~\ref{fig:XRD1} exhibits a strongly reduced intensity of the broadened layer peak near GaAs 002, i.e. the Fe$_{3}$Si 002 reflection. This indicates that the lower Fe$_{3}$Si film is severely damaged. On the other hand the Fe$_{3}$Si 220 reflection near 22.57~$^\circ$ of the uppermost film shows an increased intensity. We note, that a complete growth of the uppermost Fe$_{3}$Si at T$_S$~=~200~$^\circ$C leads to a complete vanishing of layer reflections in the corresponding XRD curve (not shown here).
Figure~\ref{fig:XRD6} shows an X-ray orientation map of the Fe$_{3}$Si 220 reflection of sample~3. A fibre texture is found in the second (upper) Fe$_{3}$Si film visible from the relatively broad maximum in the center, $\psi\leq$ 10$^\circ$. The diffracted intensity is plotted here on a logarithmic scale. Near $\psi$ = 45$^\circ$ peaks from the substrate and the first Fe$_{3}$Si film are observed exhibiting the well-known fourfold symmetry of the (001)-oriented cubic single crystal surface.
Figure~\ref{fig:TEM3}~(a) shows a cross-section HR~TEM micrograph of sample~3. The GaAs/Fe$_{3}$Si IF is still smooth, however the Fe$_{3}$Si/Al IF is very rough. The Al film and the upper Fe$_{3}$Si layer are poly-crystalline. Figure~\ref{fig:TEM3}~(b) depicts a cross-section HR~TEM micrograph of sample~3 at higher magnification. The corresponding SAD pattern is given in Fig.~\ref{fig:TEM3}~(c). There we find Fe$_{3}$Si (220) planes parallel to the GaAs surface and Fe$_{3}$Si (400) planes perpendicular to the GaAs surface and parallel to the (220) planes of the substrate. The substrate reflections are marked by open circles. The indexed fundamental reflections of Fe$_{3}$Si are dominating the SAD pattern. The Al film below is poly-crystalline without obvious preferred orientation. Probably the grains are so small that the contribution of the Al-film of sample~3 to the XRD pattern is not detected in our measurement.
Figure~\ref{fig:ebsd} depicts a SEM EBSD micrograph of sample~3 illustrating the near surface orientation of the grains of the upper Fe$_{3}$Si film. Some of the grains have a \{101\} orientation perpendicular to the substrate surface and are depicted in green. Others (shown in red) are oriented approximately along \{001\}. In addition the grains differ in the azimuthal orientation (numbered); e.~g. the azimuthal angles of rotation about an axis perpendicular to the substrate surface for the grains (1,~2,~3,~4, and ~5) are (95.2~$^\circ$, 56.9~$^\circ$, 333.2~$^\circ$, 150.2~$^\circ$, and 76.1~$^\circ$) respectively. The grain sizes are near 100~nm. Sample~3 clearly does not fulfil the criteria for device fabrication.
At the same time we  want to stress, that sample~2 is well suited for STT device preparation. Indeed magnetic tunneling devices very often are fabricated by magnetron sputtering and these films are not single crystalline.\cite{Miao2011, Walter2011, Gareev2015} Moreover, as the uppermost thicker layer acts as the fixed magnetic layer and the lower as the free switching layer, the structural quality of the interfaces as well as of the lower layer, which is a single crystal, are most important.

\section{Summary}
We have grown successfully Fe$_{3}$Si/Al/Fe$_{3}$Si multilayer systems on GaAs(001) suitable for spin-transfer torque switching of semiconductor spintronic devices. The samples were subsequently analyzed by electron microscopy and x-ray diffraction. First, only an Al film was grown on top of the epitaxial Fe$_{3}$Si/GaAs(001) structure (sample~1) and then a second Fe$_{3}$Si film was grown at low substrate temperature on top of the Al (sample~2) for comparison. Surprisingly the Al-film grows in the (111) orientation. The growth in this orientation is triggered by a thin transition region formed at the Fe$_{3}$Si/Al interface. The structural properties of sample~2 closely resemble those of sample~1. The low growth temperature range T$_s$~=~17-67~$^\circ$C of the second Fe$_{3}$Si film is therefore suitable for the preparation of spin-transfer torque  devices.  The orientation of the upper Fe$_{3}$Si film differs from that of the lower one, i. e. the Al layer and the subsequent Fe$_{3}$Si film do not grow epitaxially and exhibit a poly-crystalline structure. An approximately 2-nm-thick transition layer was detected by the different methods between the lower Fe$_{3}$Si and the Al.

\section{Acknowledgement}
The authors thank Achim Trampert and Esperanza Luna for valuable support and helpful discussion, Doreen Steffen for sample preparation, Astrid Pfeiffer
for help in the laboratory, and Claudia Herrmann  for her support during the
MBE growth, and Xiang Kong for critical reading of the manuscript.

%

\section{Tables}

\begin{table*}[htbp]
  \caption{Nominal (measured) film thicknesses, and substrate temperatures T$_S$ during epitaxial growth for three samples investigated.}

    \begin{tabular}{l c c c c c c}
    \toprule
          & sample & No.~1 & sample & No.~2 & sample & No.~3  \\
          &    thickness    &    T$_S$    &  thickness    &     T$_S$ &  thickness    &     T$_S$  \\

          & (nm)  & $^\circ$C & (nm) & $^\circ$C & (nm) & $^\circ$C \\
    \hline
    \multicolumn{1}{c}{GaAs} & 300    & 580 & 300 & 580 & 300 & 580  \\
    \multicolumn{1}{c}{Fe$_{3}$Si} & 4.5 (5.4)    & 200 & 4.5 & 200 & 4.5 & 200  \\
    \multicolumn{1}{c}{Al} & 23 (21)   & 0 & 23 (20) & 0 & 23 & 0   \\
    \multicolumn{1}{c}{Fe$_{3}$Si} & 0    & - & 45 & 17-67 & 4.5/40.5 & 13-25/200  \\
        \toprule
    \end{tabular}%
  \label{tab:tab1}%
\end{table*}%

\end{document}